# Lagrangian Description for Particle Interpretations of Quantum Mechanics – Entangled Many-Particle Case

Roderick I. Sutherland

Centre for Time, University of Sydney, NSW 2006 Australia

rod.sutherland@sydney.edu.au

A Lagrangian formulation is constructed for particle interpretations of quantum mechanics, a well-known example of such an interpretation being the Bohm model. The advantages of such a description are that the equations for particle motion, field evolution and conservation laws can all be deduced from a single Lagrangian density expression. The formalism presented is Lorentz invariant. This paper follows on from a previous one which was limited to the single-particle case. The present paper treats the more general case of many particles in an entangled state. It is found that describing more than one particle while maintaining a relativistic description requires the specification of final boundary conditions as well as the usual initial ones, with the experimenter's controllable choice of the final conditions thereby exerting a backwards-in-time influence. This retrocausality then allows an important theoretical step forward to be made, namely that it becomes possible to dispense with the usual, many-dimensional description in configuration space and instead revert to a description in spacetime using separate, single-particle wavefunctions.

## 1. Introduction

This paper focuses on interpretations of quantum mechanics in which the underlying reality is taken to consist of particles have definite trajectories at all times (e.g., [1,2]). It then enriches the associated formalism of such interpretations by providing a Lagrangian description of the unfolding events. The convenience and utility of a Lagrangian formulation is well-known from classical mechanics. The particle equation of motion, the field equation, the conserved current, action-reaction, the energy-momentum tensor, etc., are all easily derivable in a self-consistent way from a single expression. These advantages continue in the present context. Since a Lagrangian description is available in all other areas of physics and continues to be useful in modern domains such as quantum field theory and the standard model, it is appropriate to expect such a description to be relevant and applicable here as well[1].

In addition to the advantages already listed, the Lagrangian approach pursued here to describe particle trajectories also entails the natural inclusion of an accompanying field to influence the particle's motion away from classical mechanics and reproduce the correct quantum predictions. In so doing, it is in fact providing a possible explanation for why quantum behaviour exists at all – in the general case considered here, the particle is seen to be the source of a field which in turn alters the particle's trajectory via self-interaction.

---

[1] The importance of a Lagrangian approach in this context has also been emphasised by Wharton [3,4].

In a previous paper [5], the special case of a single particle was considered and a Lagrangian description for particle trajectories between measurements was tentatively proposed. The purpose of the present paper is to extend this description to the many-particle case involving entangled states. In setting up the formulation in the previous paper, the key inputs were the single-particle wavefunction $\psi(\mathbf{x},t)$, at position $\mathbf{x}$ and time t, and the associated 4-current density $j^\alpha(\mathbf{x};t)$ $(\alpha=0,1,2,3)$, the latter being a function of $\psi$. Since these quantities are both defined in four-dimensional spacetime, there was no problem in using them to construct a Lagrangian density expression which also resided in spacetime. In the many-particle case with entangled states, however, the Lagrangian density should still be defined in spacetime even though both the wavefunction of the system and the associated 4-current density expression now reside in a many-dimensional configuration space. An important result in this paper is the demonstration of a method for dealing with this mismatch successfully by replacing the configuration space description with one in spacetime[2].

The discussion will be limited to the case where the various particles involved are not presently interacting, although it will encompass entanglement existing from past interactions. The view will be taken that the case of continuing mutual interaction should be treated via quantum field theory using Feynman diagrams and this is intended to be the focus of a subsequent paper.

A key assumption which is maintained throughout this work is that Lorentz invariance remains valid. This places a strong constraint on the possible form of any particle model. It also suggests the need to include retrocausality. Taking this step provides the already-mentioned benefit of avoiding a configuration space ontology as will be discussed in Sec. 3. It also provides an escape from the various quantum no-go theorems such as those of Bell [7], Kochen and Specker [8] and Pusey, Barrett and Rudolph [9] which implicitly assume strict forwards-in-time causality.

The frequent reference to measurements in this work is not intended to imply that measurement interactions have any special status compared with other interactions[3], but merely to simplify the presentation and limit the discussion to the physical reality existing between observations. The relevant mathematics will initially be developed for the two-particle case, since the results are then directly generalisable to n particles. It will also be assumed that there is no externally applied potential. Although the mathematical formalism will be developed in Lorentz invariant form, some of the early discussion will be non-relativistic for simplicity. The formalism presented here has the advantage of being applicable generally to any wave equation of quantum mechanics. The units will be chosen such that $\hbar = c = 1$.

---

[2] Quoting David Bohm [6]: "While our theory can be extended formally in a logically consistent way by introducing the concept of a wave in 3n-dimensional space, it is evident that this procedure is not really acceptable in a physical theory, and should at least be regarded as an artifice that one uses provisionally until one obtains a better theory in which everything is expressed once more in ordinary three-dimensional space."

[3] On this point, a version of the Bohm theory of measurement applicable to the present model is given in Sec. 13 of [10].

## 2. Requirements

The Lagrangian density expression formulated in the previous paper [5] to deal with the case of a single particle will be reintroduced and discussed later in Sec. 12. It needs to be mentioned at the outset, however, that a necessary ingredient in constructing such a Lagrangian expression was found to be the inclusion of expressions for both the probability density and the current density corresponding to the particular wave equation under consideration. Now, in the non-relativistic n-particle case the standard formalism of quantum mechanics in general only provides expressions defined in 3n-dimensional configuration space for these quantities. In contrast, a satisfactory Lagrangian density for the n-particle case should continue to be defined in physical space of three spatial dimensions plus time. In the context of the present model this will be seen to entail that a separate probability density is needed for each particle. In addition, the individual probability densities need to be consistent with any statistical correlations existing in the single n-particle probability density expression $\rho(\mathbf{x}, \mathbf{x}', \mathbf{x}'', ...)$. It may seem that it is impossible to satisfy these two conditions together, since the existence of independent probability densities automatically seems to exclude correlations. Nevertheless, this apparent contradiction will be resolved easily in Sec. 8 once final boundary conditions have been incorporated and their relevance appreciated.

Similarly, it will be seen that a separate current density in spacetime is needed for each particle and that these individual currents densities each need to satisfy the continuity equation so as to conserve probability.

## 3. Entangled States

An indication of the way to proceed can be seen by considering the well-known case of a pair of particles which have previously interacted with each other and are now far apart (such as in the usual set-up employed in formulating Bell's theorem). The position coordinates of the 1st and 2nd particles will be represented by $\mathbf{x}$ and $\mathbf{x}'$, respectively. In the non-relativistic case, before any measurement is performed on either particle, the pair is represented by a single entangled wavefunction $\psi(\mathbf{x}, \mathbf{x}'; t)$. Suppose a measurement is now performed on the 1st particle and this yields a state represented by the wavefunction $\psi_1(\mathbf{x}; t)$. The 2nd particle's state can, by means of this information, be updated to a separate wavefunction $\psi_2(\mathbf{x}'; t)$. The actual form of this new wavefunction can be obtained by taking the scalar product of the 1st particle's new state $\psi_1(\mathbf{x}; t)$ with the original wavefunction $\psi(\mathbf{x}, \mathbf{x}'; t)$ as follows:

$$\psi_2(\mathbf{x}';t) = \frac{1}{N} \int_{-\infty}^{+\infty} \psi_1^*(\mathbf{x};t) \psi(\mathbf{x},\mathbf{x}';t) \, d^3x \qquad (1)$$

where N is a normalisation constant.

An important issue now arises. The collapse of the 2$^{nd}$ particle's wavefunction from the initial entangled state $\psi(\mathbf{x},\mathbf{x}';t)$ to the reduced state $\psi_2(\mathbf{x}';t)$ is meant to occur simultaneously with the measurement on the 1$^{st}$ particle. In going to the relativistic case, however, the interval between these two well-separated events is spacelike and an ambiguity arises in defining "simultaneous". Also, at whatever time the collapse is taken to occur, the introduction of Lorentz invariance means there will be a range of reference frames available in which the collapse occurs **before** the measurement. This means some sort of backwards-in-time effect, or retrocausality, is unavoidable in the standard picture as long as special relativity is assumed valid. Having heeded this point, it will now be shown how the explicit inclusion of this notion in the present formulation will enable the configuration space problem to be overcome.

Since the aim here is to consider the physical reality existing between measurements, it will be assumed that a measurement of some sort is subsequently performed on each particle. In order to satisfy the requirements specified in Sec. 2, there is a need to identify individual wavefunctions which could validly be used to describe the two particles separately during the period from when they initially move apart until when these measurements are eventually performed. At this point it is more appropriate to rewrite Eq. (1) in Lorentz invariant form and switch to Dirac notation, with the final measurement outcomes being written as $|f\rangle$ and $|f'\rangle$, respectively. Then, using the notation $x \equiv (\mathbf{x};t)$ and starting with an initial entangled state of the form $\psi(x,x') = \langle x,x'|i\rangle$, Eq. (1) is replaced by[4]:

$$\langle x'|i\rangle = \frac{1}{N} \int_{-\infty}^{+\infty} \langle f|x\rangle \hat{j}^0 \langle x,x'|i\rangle \, d^3x \qquad (2)$$

As before, this equation provides the updated wavefunction $\langle x'|i\rangle$ for the 2$^{nd}$ particle once a measurement is performed on the 1$^{st}$ particle and the result is known. Also as before, the time at which this wavefunction should become applicable is unclear because of the relativity of simultaneity. Identifying a particular spacetime hyperplane on which the change occurs is not consistent with maintaining Lorentz invariance. This being the case, it is natural to take the following extra step. It will be assumed that the updated wavefunction can be applied validly from the original time of separation of the two particles, rather than assuming it becomes applicable at some random later time. This provides the separate wavefunction needed.

---

[4] Further discussion concerning relativistic entangled states is provided in Appendix 1. Note that the operator $\hat{j}^0$ representing the zeroth component of the relevant 4-current density needs to be included in the transition to the relativistic case (Dirac or Klein-Gordon). In Eq. (2) it acts on the unprimed coordinates.

Hence there is now a separate initial wavefunction $\langle x'|i\rangle$ for the 2nd particle, taken to be valid from the moment when the two particles move apart. Eq. (2) shows clearly that this wavefunction is dependent on the final boundary condition f of the other particle, indicating the involvement of a retrocausal effect. The measurement eventually performed on the 2nd particle will provide a "final wavefunction" $\langle x'|f'\rangle$ for this particle as well. Returning to the 1st particle, an analogous procedure will provide this particle with a separate initial wavefunction $\langle x|i\rangle$ and a separate final wavefunction $\langle x|f\rangle$ for describing its state at times between the i and f events. These wavefunctions will form the basis for constructing a Lagrangian description[5]. The extension of these considerations to n particles is provided in Sec. 9.

It is important to note that the retrocausality appearing here simply corresponds to introducing final boundary conditions in addition to the usual initial ones and assuming that the choice made by the observer for either of these sets of conditions has an influence on the particle at intermediate times[6]. Two questions now arise. First, is the introduction of a separate initial wavefunction for each particle compatible with the correlated statistics predicted by the usual configuration space wavefunction $\psi(x, x')$ for the two-particle system? Second, is the state of each particle at the present time dependent on the particle's own final boundary condition as well as on the final boundary condition of the other particle? These two questions will be found to be related because, as will be shown later in Sec. 8, an affirmative answer to the first question is obtained once the second point holds true.

## 4. Reconsideration of the Single-Particle Case

Since the discussion in Sec. 8 will require the involvement of final boundary conditions even in the single-particle case, there is now a need to revisit this case to consider how such boundary conditions might be included in the formalism via some small adjustment. The situation will be made clearer by continuing with Dirac notation instead of wavefunction notation.

In a particle interpretation of quantum mechanics, the particle's position is generally not known precisely and so its future progress needs to be described in terms of a probability 4-current density. For a particle in an initial state $|i\rangle$, the general expression giving the particle's 4-current density for position **x** at time t in both the relativistic and non-relativistic cases is:

---

[5] The procedure followed in this section also has the advantage of avoiding any need for nonlocal communication between the particles at the time of measurement. The retrocausal effect is now timelike and is conveyed back via the 1st particle's final wavefunction.

[6] To be precise, the experimenter chooses the observable quantity to be measured and hence restricts the set of eigenstates which can arise. For example, the choice between two non-commuting observables for the final measurement restricts the corresponding final wavefunction to be in one of two distinct sets.

$$j^\alpha(\mathbf{x};t) = \langle i | \mathbf{x};t \rangle \hat{j}^\alpha \langle \mathbf{x};t | i \rangle \qquad (\alpha = 0,1,2,3) \qquad (3)$$

where $\hat{j}^\alpha$ is the 4-current density operator provided by the particular wave equation under consideration. In the Schrodinger case the $\hat{j}^0$ operator is simply equal to 1. Hence the expression in Eq. (3) reduces for $\alpha = 0$ to the usual probability density for the particle's position:

$$\rho(\mathbf{x};t) = \langle i | \mathbf{x};t \rangle \langle \mathbf{x};t | i \rangle \qquad (4)$$

this being the zeroth component of the Schrodinger 4-current density[7].

The state $|i\rangle$ can be thought of as the initial boundary condition imposed on the particle at some earlier time $t_i$. To avoid confusion with the time t associated with the current $j^\alpha$, Eq. (3) can be written more explicitly in the form:

$$j^\alpha(\mathbf{x};t) = \langle i;t_i | \mathbf{x};t \rangle \hat{j}^\alpha \langle \mathbf{x};t | i;t_i \rangle \qquad (5)$$

Here $\langle \mathbf{x};t | i;t_i \rangle$ is seen to be the amplitude of the state i at initial time $t_i$ onto the position state $\mathbf{x}$ at later time t. Eq. (5) then provides the conditional value of $j^\alpha(\mathbf{x};t)$ given the initial state i.

Looking at the form of Eq. (5), the obvious way to include a final boundary condition is to replace one of the initial states with a final state $|f\rangle$ relating to some later time $t_f$, as follows:

$$j^\alpha(\mathbf{x};t) = \langle f;t_f | \mathbf{x};t \rangle \hat{j}^\alpha \langle \mathbf{x};t | i;t_i \rangle \qquad (6)$$

Here $\langle \mathbf{x};t | f;t_f \rangle$ is then the amplitude of the state f at final time $t_f$ onto the position state $\mathbf{x}$ at earlier time t. Eq. (6) can be written more neatly by again using the notation $x \equiv (\mathbf{x};t)$ and deleting explicit mention of $t_i$ and $t_f$ to give:

---

[7] More generally, the relevant operators are:

Schrodinger: $\quad \hat{j}^0 = 1, \quad \hat{j}^k = \frac{1}{2im}\overleftrightarrow{\partial}^k \qquad (k = 1,2,3)$

Klein-Gordon: $\quad \hat{j}^\alpha = \frac{1}{2im}\overleftrightarrow{\partial}^\alpha \qquad (\alpha = 0,1,2,3)$

Dirac: $\quad \hat{j}^\alpha = \gamma^\alpha$

where the metric signature is $(+---)$, the operator $\overleftrightarrow{\partial}^\alpha$ stands for $\overrightarrow{\partial}^\alpha - \overleftarrow{\partial}^\alpha$ and $\gamma^\alpha$ are the Dirac matrices. In each of these cases, the resulting 4-current density is real.

$$j^\alpha(x) = \langle f|x\rangle \hat{j}^\alpha \langle x|i\rangle \tag{7}$$

The intention is for this new expression to be interpreted as giving a more specific valuation for the 4-current density once both the initial and final boundary conditions are known. The question is now whether or not this expression is able to satisfy the requirements listed earlier in Sec. 2. Subject to some minor adjustments, it will be found that expression (7) is, in fact, fully suitable. The adjustments needed are that: (i) a normalisation constant given by $\langle f|i\rangle$ needs to be included and (ii) the expression needs to be real and be fully symmetric with respect to the states i and f, both of which can be achieved simply by taking the real part. The result is:

$$j^\alpha(x) = \operatorname{Re} \frac{\langle f|x\rangle \hat{j}^\alpha \langle x|i\rangle}{\langle f|i\rangle} \tag{8}$$

This expression then describes the conditional value of $j^\alpha(x)$ given both the initial state i and the final state f. Note that this conditional wording automatically excludes the unwelcome possibility of the denominator in (8) being zero, since a zero value for the probability $|\langle f|i\rangle|^2$ would contradict the stipulation that i and f are actually found to occur.

Eq. (8) looks similar in structure to the formalism of other authors, e.g., the "weak values" of Aharonov, Albert and Vaidman [11], although the physical interpretation is different. Here it is being postulated that Eq. (8) represents the distribution of underlying trajectories in an ensemble of particles between any two successive measurements[8]. This equation summarises the extension to quantum mechanics being introduced here to resolve the many-particle case. It is consistent with the standard theory and only constitutes an "add-on", not an alteration, as will become clear in what follows. The essential point is that the final state is not generally known and averaging over it simply leads back to the standard formalism of quantum mechanics, as shown in Sec. 5.

Note that the initial and final boundary conditions are specified as Hilbert space vectors $|i\rangle$ and $|f\rangle$. In standard quantum mechanics the initial state $|i\rangle$ is the result of a measurement, or of state preparation, at time $t_i$ and summarises the particle's relevant past. By symmetry, the interpretation of state $|f\rangle$ is taken to be similar. Here it will be assumed for convenience to be the result of the next measurement performed on the particle[9], this being performed at some later time $t_f$. This result is taken to summarise the particle's relevant future. Also note that the position **x** in Eq. (8) is not a measurable quantity. The only measurement outcomes are i and f. In contrast, **x** represents the hidden position of the particle at times between these two measurements. It should be mentioned here that expression (8) conserves probability flow, as will be highlighted in Sec. 18. Finally, for the purposes of the proof in the next section note that, since f is a measurement result, it will correspond to one of the eigenstates of a complete orthonormal set.

---

[8] In this context it is also necessary that $j^\alpha(x)$ be real, since a complex distribution is not ontologically meaningful.
[9] As mentioned in the introduction, the frequent reference to measurements here is not suggesting that measurement interactions have any special status compared with other interactions, but merely aiming to simplify the discussion.

## 5. Consistency with the Standard Current Density of Quantum Mechanics

If the conditional dependence of the current on both i and f is included explicitly in the notation, the left hand side of Eq. (8) should be written $j^\alpha(x|i,f)$. In contrast, the standard 4-current density (3) predicted by quantum mechanics should be written $j^\alpha(x|i)$, since it depends only on the initial state. Using this notation, it is easily seen that expression (8) reduces to the standard 4-current expression when a weighted average is taken over the unknown final state. The two 4-current densities are expected to be related via:

$$j^\alpha(x|i) = \int_{-\infty}^{+\infty} j^\alpha(x|i,f) \rho(f) df \qquad (9)$$

where $\rho(f)$ is the probability density for the subsequent measurement result f. Inserting the usual expression:

$$\rho(f) = |\langle f|i\rangle|^2 \qquad (10)$$

together with (8) into Eq. (9) yields:

$$\begin{aligned} j^\alpha(x|i) &= \int_{-\infty}^{+\infty} \operatorname{Re} \frac{\langle f|x\rangle \hat{j}^\alpha \langle x|i\rangle}{\langle f|i\rangle} |\langle f|i\rangle|^2 df \\ &= \operatorname{Re} \int_{-\infty}^{+\infty} \langle i|f\rangle \langle f|x\rangle \hat{j}^\alpha \langle x|i\rangle df \\ &= \operatorname{Re} \left[ \langle i|x\rangle \hat{j}^\alpha \langle x|i\rangle \right] \end{aligned}$$

(11)

This result is then seen to be equivalent to the standard 4-current density (3) as required, since the factor in the square bracket is always real as mentioned in footnote 7.

## 6. Hidden Trajectories Between Measurements

The time (or zeroth) component of the 4-current density vector in Eq. (8) alternates between positive and negative values. This means that the 4-current lines in spacetime must have sections which point backwards in time, in addition to the usual forwards-in-time parts (see, e.g., Fig. 2 in [10]). These lines also curve continuously and smoothly, which means the current density 4-vector must pass through spacelike directions as well. This may seem surprising but it should be remembered that, under a particle interpretation, the standard 4-current density provided by the Klein-Gordon equation has the same properties. Note that the 4-current lines of the new $j^\alpha$ expression are covariant, as required. Furthermore, there is no conflict with experiment in the present circumstances because the form of expression (8) indicates that any such "transluminal" behaviour occurs between measurements and is never actually observed[10]. Such an observation would require a position measurement whereas, as (8) shows, the next measurement is generally of a different variable f. In these circumstances, the fact that the time component of the 4-current density is not positive definite is not important because it is only describing the direction of this 4-vector in spacetime and need not be used for probability predictions. For convenience of terminology, however, this component will continue to be referred to here as the "probability density", even though it is not restricted to positive values. In any case, it should be remembered that this component is always positive in the local rest frame [12] of the 4-current density even though not always in our own frame. Hence it can be interpreted as a probability density in the local rest frame.

In the special case where the next measurement is actually one of position, it is sufficient for the soundness of the model if it can be demonstrated that the time component of the 4-current density becomes positive as the measurement time approaches (so that the world line then definitely points forwards in time) and that it becomes zero for all position values other than the one measured. It is shown in Appendix 2 that these two requirements do, in fact, hold true.

### 7. Separate 4-Current Densities

The entangled pair of particles discussed in Sec. 3 will now be revisited. Having introduced individual wavefunctions for the two particles, a separate 4-current density for each particle can now be obtained, these expressions being defined in spacetime rather than in configuration space[11]. For example, using the 1st particle's individual wavefunctions $\langle x|i\rangle$ and $\langle x|f\rangle$, this particle's 4-current density is simply given by the expression in Eq. (8):

$$j_1^\alpha(x) = \text{Re}\frac{\langle f|x\rangle \hat{j}^\alpha \langle x|i\rangle}{\langle f|i\rangle} \tag{12}$$

i.e., the same expression as when there is only one particle present rather than a pair. By changing to wavefunction notation:

$$\psi_i(x) \equiv \langle x|i\rangle \tag{13}$$

$$\psi_f(x) \equiv \langle x|f\rangle \tag{14}$$

the 1st particle's 4-current density can also be written in the form:

$$j_1^\alpha(x) = \text{Re}\frac{\psi_f^*(x) \hat{j}^\alpha \psi_i(x)}{\langle f|i\rangle} \tag{15}$$

---

[10] Nevertheless it is instructive to compare the (hidden) pictures in three dimensions and four dimensions for a world line that gradually doubles back in time and then bends smoothly forwards again in this way. From a four-dimensional viewpoint, the picture consists of just a single, s-shaped curve in spacetime. In terms of time evolution in three dimensions, however, the picture consists of the creation and then annihilation of particle-antiparticle pairs. Since the smooth world lines being considered here differ from Feynman zigzags in having no sharp corners, the particle-antiparticle description has the disadvantage of not being Lorentz invariant in this case. This lack of invariance of the three-dimensional picture is due to the fact that the points at which a smooth world line reverses its time direction are actually frame dependent, so that different observers would not agree in specifying the precise spacetime event at which creation or annihilation occurs. For this reason, the four-dimensional picture will be preferred here over the three-dimensional description. Further discussion of transluminal motion is provided in Appendix 3.

Here, in accordance with the discussion in previous sections, $\psi_i(x)$ is the particle's initial wavefunction at time t, summarising the initial boundary condition[12] imposed on this particle at an earlier time $t_i$, and $\psi_f(x)$ is its final wavefunction at the same time t, summarising the final boundary condition imposed at a later time $t_f$. Note that the wavefunctions simply transport the boundary conditions to a more convenient time. A 4-current density expression for the 2nd particle can be obtained analogously by combining that particle's separate initial and final wavefunctions.

## 8. Demonstration of Consistency Between Separate Probability Densities and Observed Correlations

It is now necessary to demonstrate that the separate expressions described in the previous section are consistent with the known, correlated statistics obtained when position measurements are performed on the two particles. For simplicity, this analysis will be carried out for the non-relativistic case[13]. It is useful at this point to introduce the joint probability density[14] $\rho(x_f, x'_f)$ for the 1st particle to be at position $\mathbf{x}_f$ at time $t_f$ and the 2nd particle to be at $\mathbf{x}'_f$ at time $t'_f$. Now, as stated in Sec. 4, the present model simply assumes the standard predictions of quantum mechanics and then adds a small extension to the formalism to include final boundary conditions. Therefore it is a basic postulate of the model that the probability density $\rho(x_f, x'_f)$ can still be identified with the usual expression predicted by quantum mechanics, as follows:

$$\rho(x_f, x'_f) = |\langle x_f, x'_f | i \rangle|^2 \qquad (16)$$

---

[11] This constitutes an advantage over the weak values theory mentioned in Sec. 4, in which it is not possible to split up the configuration space wavefunction of the many-particle case into separate single-particle wavefunctions.

[12] For completeness it should be mentioned here that, as shown by Eq. (2), each particle's initial wavefunction may depend not only on the initial boundary condition for that particle but also on the final boundary conditions of other particles with which it has interacted.

[13] In the relativistic case there is the well-known complication in standard quantum mechanics (see, e.g., Sec. 3c in [13]) that the usual position variable **x** to be discussed here is not actually an observable, since its eigenstates are not orthogonal and its operator is not Hermitian.

The task here then is to show that this relationship is consistent at the time of measurement with the separate expressions introduced in Sec. 7. Returning to Eq. (12) in that section, the non-relativistic probability density for the 1st particle to be at position $\mathbf{x}$ at intermediate time t given that its subsequent final position is measured to be $\mathbf{x}_f$ at time $t_f$ is:

$$\rho_1(x | x_f) = \mathrm{Re} \frac{\langle x_f | x \rangle \langle x | i \rangle}{\langle x_f | i \rangle} \qquad (17)$$

This expression is also conditional on the initial state i (as is the joint probability density $\rho(x_f, x'_f)$ above), but for simplicity this will not be included in the left hand side notation. Similarly, the 2nd particle's probability density for position $\mathbf{x}'$ at time $t'$ given final position $\mathbf{x}'_f$ at time $t'_f$ is:

$$\rho_2(x' | x'_f) = \mathrm{Re} \frac{\langle x'_f | x' \rangle \langle x' | i \rangle}{\langle x'_f | i \rangle} \qquad (18)$$

Combining these two conditional probabilities yields the joint probability density for x and x′ given both $x_f$ and $x'_f$:

$$\rho(x, x' | x_f, x'_f) = \rho_1(x | x_f) \rho_2(x' | x'_f) \qquad (19)$$

The simple product involved here indicates that there are no correlations at this stage, which superficially suggests that Eq. (19) is not compatible with the correlations inherent in Eq. (16). Such a conclusion would be wrong, however, because we are not yet comparing like with like. It needs to be remembered that the uncorrelated form (19) is conditional on the final positions $x_f$ and $x'_f$. It assumes these future results are both known, which is not the normal situation. Correlations will, in fact, arise once a weighted average is taken over all the possible, unknown values for $x_f$ and $x'_f$. To obtain this average, it is first necessary to construct the joint probability distribution for all four variables x, x′, $x_f$ and $x'_f$ together:

---

[14] This probability density is defined as per the discussion in Appendix 1.

$$\rho(x, x', x_f, x'_f) = \rho_2(x, x'|x_f, x'_f)\rho(x_f, x'_f) \tag{20}$$

In doing this, it is permissible and in accordance with the model to employ both (16) and (19) together. Inserting these equations into Eq. (20) yields:

$$\rho(x, x', x_f, x'_f) = \rho_1(x|x_f)\rho_2(x'|x'_f)|\langle x_f, x'_f|i\rangle|^2 \tag{21}$$

The probability density for just $x$ and $x'$ is then found by integrating over all $x_f$ and $x'_f$:

$$\rho(x, x') = \int_{-\infty}^{+\infty}\int_{-\infty}^{+\infty} \rho_1(x|x_f)\rho_2(x'|x'_f)|\langle x_f, x'_f|i\rangle|^2 d^3x_f d^3x'_f \tag{22}$$

This is no longer a simple product and so correlations have appeared, as required, in the process of eliminating the unknown final states. Hence the possibility of reconciling correlations with the use of separate wavefunctions has been successfully demonstrated.

A further point now needs to be established in order to ensure consistency. At this stage it is not important whether the correlations in Eq. (22) are the correct ones or not because they refer to the hidden positions $\mathbf{x}$ and $\mathbf{x}'$ at intermediate times $t$ and $t'$, respectively. It is necessary, however, to show that the correct correlations will emerge in agreement with Eq. (16) as the times of the actual observations are approached via $t \to t_f$ and $t' \to t'_f$, since this is when consistency with the standard predictions is needed. For this purpose, the expressions for $\rho_1(x|x_f)$ and $\rho_2(x'|x'_f)$ given in Eqs. (17) and (18) will now be examined[15].

Eq. (17) can be written out with all times shown explicitly as follows:

$$\rho_1(\mathbf{x}, t|\mathbf{x}_f, t_f) = \mathrm{Re}\frac{\langle \mathbf{x}_f, t_f|\mathbf{x}, t\rangle \langle \mathbf{x}, t|i\rangle}{\langle \mathbf{x}_f, t_f|i\rangle} \tag{23}$$

In approaching the limit $t \to t_f$, it is a standard result that the term $\langle \mathbf{x}_f, t_f|\mathbf{x}, t\rangle$ gradually becomes a delta function:

$$\langle \mathbf{x}_f, t_f|\mathbf{x}, t\rangle \to \delta^3(\mathbf{x} - \mathbf{x}_f) \tag{24}$$

Hence in this limit Eq. (23) becomes:

$$\rho_1(\mathbf{x}, t_f|\mathbf{x}_f, t_f) = \mathrm{Re}\frac{\delta^3(\mathbf{x} - \mathbf{x}_f)\langle \mathbf{x}, t_f|i\rangle}{\langle \mathbf{x}_f, t_f|i\rangle} \tag{25}$$

The delta function then allows cancellation to yield:

---

[15] The next few steps involve repeating the procedure followed in Eqs. (90) to (93) of Appendix 2. They will be included here as well to make the present proof self-contained.

$$\rho_1(\mathbf{x},t_f|\mathbf{x}_f,t_f) = \delta^3(\mathbf{x}-\mathbf{x}_f) \tag{26}$$

Similarly, taking the limit $t' \to t'_f$ in the 2$^{nd}$ particle's probability density $\rho_2(x'|x'_f)$ yields:

$$\rho_2(\mathbf{x}',t'_f|\mathbf{x}'_f,t'_f) = \delta^3(\mathbf{x}'-\mathbf{x}'_f) \tag{27}$$

Finally, substituting the results (26) and (27) into Eq. (22) then gives:

$$\lim_{\substack{t \to t_f \\ t' \to t'_f}} \rho(x,x') = \int_{-\infty}^{+\infty}\int_{-\infty}^{+\infty} \delta^3(\mathbf{x}-\mathbf{x}_f)\delta^3(\mathbf{x}'-\mathbf{x}'_f)|\langle \mathbf{x}_f,t_f;\mathbf{x}'_f,t'_f|i\rangle|^2 \, d^3x_f \, d^3x'_f \tag{28}$$

which reduces on integration to:

$$\lim_{\substack{t \to t_f \\ t' \to t'_f}} \rho(x,x') = |\langle \mathbf{x},t_f;\mathbf{x}',t'_f|i\rangle|^2 \tag{29}$$

This result for the limit is in accordance with the standard quantum mechanical prediction (16), as required, and hence contains the correct correlations between the measured positions.

The above discussion highlights the fact that expression (29) arises via a gradual and continuous process as the measurement time grows near. This is seen to be a simple consequence of the fact that the final boundary conditions $|\mathbf{x}_f,t_f\rangle$ and $|\mathbf{x}'_f,t'_f\rangle$ are exerting influences via the standard quantum mechanical amplitudes $\langle \mathbf{x},t|\mathbf{x}_f,t_f\rangle$ and $\langle \mathbf{x}',t'|\mathbf{x}'_f,t'_f\rangle$, respectively, and that these amplitudes reduce continuously to delta functions as t approaches $t_f$ and $t'$ approaches $t'_f$.

In summary, the apparently independent probability densities for the two particles become consistent with the known correlations once the unknown final positions have been integrated out.

## 9. The n-Particle Case

The formalism will now be extended from 2 to n particles. Consider a configuration space wavefunction $\psi_i(x,x',x'',...)$ which allows for entanglement existing from past interactions. It will be assumed that measurements are eventually performed on all the particles, with the outcomes being described by the separate wavefunctions $\psi_f(x)$, $\psi_{f'}(x')$, $\psi_{f''}(x'')$, ... . Taking the 1$^{st}$ particle as an example and working by analogy with Eq. (2), the updated initial wavefunction which becomes applicable for this particle as a result of the measurements performed on all the other particles is given by:

$$\psi_i(x) = \frac{1}{N}\int_{-\infty}^{+\infty}\int_{-\infty}^{+\infty}...\psi_{f'}(x')\psi_{f''}(x'')...\hat{j}^{0'}\hat{j}^{0''}...\psi_i(x,x',x'',...) \, d^3x' \, d^3x''... \tag{30}$$

where N is an appropriate normalisation constant. Note that the right hand side of this equation does not involve the 1st particle's measurement result $\psi_f(x)$ (all other results being included). Using the argument presented in Sec. 3, the wavefunction given by Eq. (30) is assumed to be already valid at earlier times. Hence the 1st particle can now be assigned this separate initial wavefunction $\psi_i(x)$, together with the final wavefunction $\psi_f(x)$ generated by this particle's subsequent measurement outcome. These two wavefunctions can then be combined to provide an individual 4-current density via Eq. (15).

Assigning each particle a separate pair of wavefunctions and a separate 4-current density in this way, it is now possible to use similar formalism to the single-particle case in order to describe the n-particle case as well. There will then be a separate Lagrangian density for each particle.

## 10. Lagrangian Formalism

The approach that will be pursued here is to work by analogy with the familiar classical formalism for a charged particle interacting with an electromagnetic field, where an overall Lagrangian density $\mathcal{L}$ is able to describe both the particle and the field potential together. For a single particle, this electromagnetic Lagrangian density can be broken up in the following convenient way:

$$\mathcal{L} = \mathcal{L}_{\text{field}} + \mathcal{L}_{\text{particle}} + \mathcal{L}_{\text{interaction}} \tag{31}$$

In terms of the particle's 4-velocity $u^\alpha$ and the electromagnetic 4-potential $A^\alpha(x)$, the explicit expressions for the three terms in this equation are [5]:

$$\mathcal{L} = -\tfrac{1}{4}F_{\alpha\beta}F^{\alpha\beta} - \sigma_0 m(u_\alpha u^\alpha)^{\frac{1}{2}} - \sigma_0 q\, u_\alpha A^\alpha \qquad (\alpha,\beta = 0,1,2,3) \tag{32}$$

Here $F^{\alpha\beta}$ is the electromagnetic field tensor, which can be expressed instead in terms of derivatives of $A^\alpha$, and m and q are the particle's mass and charge, respectively. The quantity $\sigma_0$ is the rest density distribution of the particle through space. This involves a delta function because, at any time, the particle's "matter density" is all concentrated at one point. The explicit form of $\sigma_0$ is:

$$\sigma_0 = \frac{1}{u^0}\delta^3[\mathbf{x} - \mathbf{x}_p(\tau)] \tag{33}$$

where $\mathbf{x}_p$ is the particle's spatial position as a function of proper time $\tau$ and $\mathbf{x}$ is an arbitrary point in space. Note that Eq. (32) is written in manifestly Lorentz invariant form. It is assumed that the metric tensor has signature $(+---)$.

The question now arises as to what form of Lagrangian density could be proposed, by analogy with the electromagnetic example, for use within a particle interpretation of quantum mechanics. As was discussed for the single-particle case, the first step is to note that the $\mathcal{L}_{field}$ term in (31) can simply be identified with one of the well-known Lagrangian densities for the $\psi$ field alone, as quoted in quantum mechanics texts. The choice depends on which wave equation is under consideration, each expression being formed from the corresponding wavefunction and its complex conjugate. The problem is then to include suitable terms for $\mathcal{L}_{particle}$ and $\mathcal{L}_{interaction}$ which will describe the particle motion as well. These terms are also expected to contain the wavefunction $\psi$ so that the particle will be guided appropriately.

## 11. Relevant Quantum Formalism

At this point, a short digression is necessary. In constructing the particle and interaction terms for the single-particle Lagrangian density in the previous paper [5], it was found useful to employ the following standard formalism which applies for any wave equation of quantum mechanics:

Each wave equation has a 4-current density expression $j^\alpha(x)$ associated with it. For example, the 4-current density for the Dirac equation is[16]:

$$j^\alpha = \bar{\psi}\gamma^\alpha\psi \tag{34}$$

The 4-current density $j^\alpha(x)$ can always be written in the form:

$$j^\alpha = \rho_0 \bar{u}^\alpha \tag{35}$$

where $\rho_0(x)$ and $\bar{u}^\alpha(x)$ are the rest density and 4-velocity of the probability flow, respectively. Note that these quantities are quite different from the particle rest density $\sigma_0$ and the particle 4-velocity $u^\alpha$. Also note that $\rho_0$ is uniquely determined once $j^\alpha$ is given, as can be seen by applying the identity $\bar{u}_\alpha \bar{u}^\alpha = 1$ as follows:

$$j_\alpha j^\alpha = (\rho_0 \bar{u}_\alpha)(\rho_0 \bar{u}^\alpha) = \rho_0^2 \tag{36}$$

and therefore:

$$\rho_0 = (j_\alpha j^\alpha)^{1/2} \tag{37}$$

This standard formalism will also be useful for the many-particle case discussed in the next section. Note that in quantum mechanics the 4-current density $j^\alpha(x)$ usually varies with position and time even in the absence of any externally applied potential, i.e., the flow lines are generally curved, not straight. This means that the motion of the individual particles presumed to underlie this probability flow must also be non-uniform, i.e., the particles seem to behave as if some sort of field is acting on them.

---

[16] Here $\bar{\psi}$ is the adjoint of $\psi$, rather than the complex conjugate.

## 12. Proposed New Lagrangian Density

Returning to Eqs. (31) and (32), a Lagrangian density analogous to these equations and suitable to use with a particle interpretation of quantum mechanics will now be presented. As shown in [5], the following expression is adequate for the single-particle case[17]:

$$\mathcal{L} = \mathcal{L}_{field} + \mathcal{L}_{particle} + \mathcal{L}_{interaction}$$
$$= \mathcal{L}_{field} - \sigma_0 \rho_0 (u_\alpha u^\alpha)^{1/2} + \sigma_0 u_\alpha j^\alpha \quad (38)$$

It is constructed simply by taking the electromagnetic expression (32) and replacing the 4-vector potential $A^\alpha$ and the mass m by, respectively, the 4-current density $j^\alpha(x)$ and the rest density $\rho_0(x)$ of standard quantum mechanics, these latter two quantities having been discussed here in the previous section[18]. A more general Lagrangian density is now needed to describe the many-particle case as well. This can be achieved by instead choosing $j^\alpha(x)$ to be the new expression developed in the present paper, rather than the standard expression. Specifically, referring to Eq. (15), the 4-current density for each particle will be chosen to be of the form:

$$j^\alpha(x) = \text{Re} \frac{\psi_f^*(x) \hat{j}^\alpha \psi_i(x)}{\langle f | i \rangle} \quad (39)$$

where the final boundary conditions are now included. The many-particle Lagrangian density is then expected to be similar to (38).

A further complication, however, is that the current density 4-vector in Eq. (39) can range over both timelike and spacelike directions as discussed in Sec. 6. Some consequences of this are discussed in detail in Appendix 4. First, the velocity identity $u_\alpha u^\alpha = 1$ changes to $u_\alpha u^\alpha = -1$ when the velocity 4-vector becomes spacelike. Second, there is a need to take the absolute value in defining the new rest density $\rho_0(x)$ for each particle. Modifying Eq. (37), it is now given in terms of the new $j^\alpha(\mathbf{x},t)$ via:

$$\rho_0 = \left| j_\alpha j^\alpha \right|^{1/2} \tag{40}$$

---

[17] In the previous paper it was noted that the second and third terms of (38) should each contain an arbitrary constant k with dimensions $[ML^3]$ to ensure that the units balance. As before, this constant will be set equal to one to keep the formalism simple. (The choice of units $\hbar = c = 1$ also hides a factor of c which would otherwise be present in the second term.)

[18] Note that $u^\alpha$ is an independent variable in this Lagrangian density and should not be confused with the probability 4-velocity $\bar{u}^\alpha = j^\alpha / \rho_0$, which is dependent on the wavefunction. Also note that $\bar{u}^\alpha$ is a function of position, whereas $u^\alpha$ is not.

Taking both these points into account, the new Lagrangian density is finally defined to have the form[19]:

$$\mathcal{L} = \mathcal{L}_{field} \mp \sigma_0 \rho_0 \left| u_\alpha u^\alpha \right|^{1/2} + \sigma_0 u_\alpha j^\alpha \tag{41}$$

where the second term is modified slightly compared with (38) and it is understood that the upper sign applies for timelike $u^\alpha$ and the lower sign applies for spacelike $u^\alpha$. Here $j^\alpha(x)$ and $\rho_0(x)$ are defined as in Eqs. (39) and (40). An additional point concerning this transition to the new Lagrangian density is that the final wavefunction $\psi_f$ needs to be included in the $\mathcal{L}_{field}$ term as well. The method of including it will be illustrated in the next two sections.

As per the discussion in Sec. 9, this approach allows each of the particles encompassed by a non-factorisable, many-particle wavefunction to be described by a separate Lagrangian density. The general Lagrangian density for n particles will then be the sum:

$$\mathcal{L} = \mathcal{L}_1 + \mathcal{L}_2 + \ldots + \mathcal{L}_n \tag{42}$$

where each term in this equation will be of the form (41).

It is important to be clear on the wavefunction dependence of the new Lagrangian expression. For the single-particle case in [5], the quantities $\mathcal{L}_{field}$, $j^\alpha$ and $\rho_0$ were simply all functions of the wavefunction $\psi_i(x)$ and its complex conjugate $\psi_i^*(x)$ (although the subscript i was not explicitly included in that work). Here, for each particle, they are now functions of $\psi_i(x)$ and $\psi_f^*(x)$ instead. Hence the final boundary conditions are introduced into the model through this change. The resulting new formalism developed here replaces the previous single-particle formalism in [5] and can be applied in both the single-particle and many-particle cases.

Some explicit examples of Lagrangian densities having the proposed form will now be presented to gain further insight.

---

[19] As mentioned in the previous footnote, $u^\alpha$ is an independent variable here. This independence will, however, be affected when the joint distribution in Eq. (54) is introduced later.

## 13. Example: Klein-Gordon Case

A specific model encompassed by the general expression (41) is the Lagrangian density for Klein-Gordon particles. In this case, each particle in an entangled many-particle state $\psi_i(x, x', x'', ...)$ will be described (in accordance with Eqs. (31) and (41)) by an individual Lagrangian density $\mathcal{L}$ of the form:

$$\mathcal{L} = \frac{1}{2m} \text{Re} \left[ \frac{1}{\langle f|i \rangle} \left( \psi_f^* \overleftrightarrow{\partial}_\alpha \overrightarrow{\partial}^\alpha \psi_i - m^2 \psi_f^* \psi_i \right) \right] \mp \sigma_0 \rho_0 \left| u_\alpha u^\alpha \right|^{1/2} + \sigma_0 u_\alpha j^\alpha \qquad (43)$$

where the single-particle wavefunction $\psi_i(x)$ in this expression is provided by the mechanism in Eq. (30) and where $j^\alpha(x)$ and $\rho_0(x)$ are defined by[20]:

$$j^\alpha = -\frac{1}{2m} \text{Re} \frac{\psi_f^* i \overleftrightarrow{\partial}^\alpha \psi_i}{\langle f|i \rangle} \qquad (44)$$

and:

$$\rho_0 = \left| j_\alpha j^\alpha \right|^{1/2} \qquad (45)$$

Note the way in which the final wavefunction has been included in the $\mathcal{L}_{\text{field}}$ term. This term is linear in both $\psi_i(x)$ and $\psi_f^*(x)$. The fact that the real part has been taken in both $\mathcal{L}_{\text{field}}$ and $j^\alpha$ means that these expressions are symmetric in i and f and has the consequence that $\psi_i^*(x)$ and $\psi_f(x)$ are implicitly present here as well.

## 14. Example: Dirac Case

The Dirac case is similar in that each particle in an entangled many-particle state will be described by an individual Lagrangian density $\mathcal{L}$ of the form:

$$\mathcal{L} = \text{Re} \left[ \frac{1}{\langle f|i \rangle} \left( -i \bar{\psi}_f \gamma^\alpha \partial_\alpha \psi_i + m \bar{\psi}_f \psi_i \right) \right] \mp \sigma_0 \rho_0 \left| u_\alpha u^\alpha \right|^{1/2} + \sigma_0 u_\alpha j^\alpha \qquad (46)$$

where now the adjoint spinor $\bar{\psi}_f$ is used instead of $\psi_f^*$ and where:

---

[20] The operator $\overleftrightarrow{\partial}^\alpha$ in Eq. (44) stands for $\overrightarrow{\partial}^\alpha - \overleftarrow{\partial}^\alpha$. Also note that the letter i is playing two roles here, representing both "initial" and $\sqrt{-1}$.

$$j^\alpha = \text{Re}\frac{\overline{\psi}_f \gamma^\alpha \psi_i}{\langle f | i \rangle} \tag{47}$$

and:

$$\rho_0 = \left| j_\alpha j^\alpha \right|^{1/2} \tag{48}$$

Again, the dependence on the final conditions via $\overline{\psi}_f(x)$ can be seen explicitly[21].

## 15. Introduction of a Statistical Element via a Joint Distribution

The Lagrangian model formulated here is not inherently statistical. Once the initial and final boundary conditions are specified, this uniquely determines the wavefunctions and hence $j^\alpha$ and $\rho_0$ as well. These, together with the initial position and velocity of each particle, then determine the particle trajectories. Quantum mechanics, on the other hand, is a statistical theory. To compare the predictions of the two theories, it is therefore necessary to introduce a statistical element into the present model. This will be achieved by assuming a lack of knowledge of each particle's position $\mathbf{x}$ and 3-velocity $\mathbf{v}$ at time t and hence by describing these two variables via a joint probability distribution $P(\mathbf{x},\mathbf{v};t)$.

It will be sufficient for the discussion in this section to focus on just the single-particle case. A condition for consistency with quantum mechanics is then that the probability density $\rho(\mathbf{x};t)$ and current density $\mathbf{j}(\mathbf{x};t)$ predicted by the relevant wave equation should be related to the chosen joint distribution $P(\mathbf{x},\mathbf{v};t)$ in the following ways[22]:

$$\int_{-\infty}^{+\infty} P(\mathbf{x},\mathbf{v};t)\, d^3v = \rho(\mathbf{x};t) \tag{49}$$

$$\int_{-\infty}^{+\infty} P(\mathbf{x},\mathbf{v};t)\, \mathbf{v}\, d^3v = \mathbf{j}(\mathbf{x};t) \tag{50}$$

These equations implicitly entail that $P(\mathbf{x},\mathbf{v};t)$ must be a function of $\psi$. A variety of different particle models are then possible by choosing different expressions for $P(\mathbf{x},\mathbf{v};t)$.

---

[21] Since, as mentioned in Sec. 1, the present discussion is limited to the free-space case, the various Lagrangian densities presented here do not contain any external potential. Terms containing such potentials can easily be added, however. For example, the appropriate extra term for including an external 4-vector potential $A^\alpha$ in the Dirac case would be of the form $A_\alpha j^\alpha$, where $j^\alpha$ is given by Eq. (47).

[22] The fact that these equations can be taken as a sufficient condition for consistency is argued in Bohm's theory of measurement [1,2]. Note that the particle velocities discussed here correspond to values existing between measurements but not necessarily to measurement outcomes (since the measurement interaction may gradually modify the state during the measurement process, as in the Bohm model).

Note that each model is expected to allow a range of different velocity values at each point in space-time (i.e., it is expected to be a "phase space" model). Progress is usually impeded at this point by the requirement that the probability density $P(\mathbf{x},\mathbf{v};t)$ should also be positive. For example, the Wigner distribution [14] is often used for practical calculations, but the fact that it is not positive means that it is not physically viable for present purposes.

In writing Eqs. (49) and (50), it has been implicitly assumed that the joint distribution introduced here is conditional on the initial boundary conditions but not the final ones. In the present context, however, it is more convenient to have a joint distribution which is conditional on both the initial and the final conditions. It is also more convenient to have the relevant equations in manifestly Lorentz invariant form. These two issues can both be resolved by re-expressing Eqs. (49) and (50) in the form of the following single equation:

$$\int_{-\infty}^{+\infty} P_0(\mathbf{x},u;t)\, u^\alpha \, d^4 u = j^\alpha(\mathbf{x};t) \qquad (\alpha = 0,1,2,3) \qquad (51)$$

where Eq. (49) corresponds to inserting $\alpha = 0$ whilst Eq. (50) corresponds to $\alpha = 1,2,3$. Here a switch has been made to 4-velocity $u^\alpha$ instead of 3-velocity $\mathbf{v}$ and it is understood that $j^\alpha(\mathbf{x};t)$ is the new 4-current density expression (39) of the present model. The function $P_0(\mathbf{x},u;t)$ in Eq. (51) is an invariant distribution. It is related to the usual joint probability distribution $P(\mathbf{x},u;t)$ via:

$$P(\mathbf{x},u;t) = P_0(\mathbf{x},u;t)\, u^0 \qquad (52)$$

where it should be kept in mind that the usual distribution is the zeroth component of a 4-vector. Both $P_0(\mathbf{x},u;t)$ and $P(\mathbf{x},u;t)$ are understood here to be conditional on both the initial and final boundary conditions $|i\rangle$ and $|f\rangle$.

In view of Eq. (51), the issue now becomes one of choosing a suitable expression for the invariant distribution $P_0(\mathbf{x},u;t)$ in order to complete the present model. In addition to satisfying (51), this expression must also be positive and be consistent with the chosen Lagrangian density. On this last point, the specific requirement is that the particle acceleration $\dfrac{du^\alpha}{d\tau}$ (an expression for which will be provided by the Lagrangian equation of motion for the particle) together with the distribution expression $P_0(\mathbf{x},u;t)$ must mutually satisfy the continuity equation in phase space:

$$\frac{\partial}{\partial u^\alpha}\left[P_0(\mathbf{x},u;t)\frac{du^\alpha}{d\tau}\right] + \frac{\partial}{\partial x^\alpha}\left[P_0(\mathbf{x},u;t)\frac{dx^\alpha}{d\tau}\right] = 0 \qquad (53)$$

Hence the aim is actually to find both an expression for $\mathscr{L}$ and an expression for $P_0(\mathbf{x}, \mathbf{u}; t)$ which are mutually consistent. Despite an extensive search, the present author has only been able to find one solution which satisfies all of these conditions. The Lagrangian density expression is, of course, the one already given in Eq. (41) and the joint distribution expression is as follows:

$$P_0(\mathbf{x}, \mathbf{u}; t) = \rho_0 \, \delta^4\left(u^\alpha - \frac{j^\alpha}{\rho_0}\right) \tag{54}$$

where $j^\alpha(\mathbf{x}; t)$ is the new 4-current density as in Eq. (39) and $\rho_0(\mathbf{x}; t)$ is the corresponding rest density defined in Eq. (40).

The delta function in this distribution effectively limits the particle's velocity to being a function of position $\mathbf{x}$ (as can be seen in Eq. (55) below). Consequently, there is not a range of possible velocities at each position but a unique value instead. This reduces the description from a phase space model to a configuration space model. At first sight this seems surprisingly restrictive on the particle's velocity. It should be noted, however, that the particle is not just moving in an independent external field. As will be discussed in more detail shortly, the Lagrangian formalism involves the particle acting as the source of the field as well. This leaves very little room to manoeuvre in identifying consistent trajectories for the particle. Hence the strong restriction on the velocity range at each position.

The presence of the delta function in the distribution (54) also has the surprising consequence that this distribution is already supplying an equation of motion for the particle without the need to obtain one from the Lagrangian formalism. The equation of motion implied by the delta function is clearly[23]:

$$u^\alpha = \frac{j^\alpha(\mathbf{x}; t)}{\rho_0(\mathbf{x}; t)} \tag{55}$$

This equation is seen to be similar in appearance to the guidance equation of the relativistic Bohm model, although it should be kept in mind that the right hand side here implicitly contains the final boundary conditions as well as the initial ones[24]. The particle velocity given by Eq. (55) refers to times between measurements and so is not directly observable. It will, however, play an important role in later sections in ensuring that other predictions of the present Lagrangian formalism are in agreement with quantum mechanics. It will also be necessary at a later point to establish the consistency of this equation of motion with the one to be supplied by the Lagrangian formalism.

---

[23] Eq. (55) constrains the particle velocity to be equal to the local current velocity at the particle's location. This equation is, of course, only applicable at locations where the particle is present - there is still a current velocity elsewhere but no particle velocity.

[24] A non-relativistic version of the retrocausal Bohm model presented here is given in [10].

Finally it should be noted that, since the formalism of the present model provides a separate 4-current density expression for each of the n particles in an entangled state, it also provides a separate 4-velocity expression via Eq. (55) for each particle as well. Each of these 4-velocities is defined in spacetime, whereas non-retrocausal theories such as the standard Bohm model can only provide an n-particle velocity expression defined in configuration space.

The remaining sections of this paper are now devoted to summarising the various consequences which flow from the general Lagrangian density (41) introduced earlier. These can be obtained by standard Lagrangian techniques. It will be confirmed that the new Lagrangian density yields predictions which are consistent with quantum mechanics.

It should also be noted that this description is providing a reason for why quantum mechanical effects exist at all. The choice of a combined particle/field Lagrangian density entails that the particle is the source of a field which influences the particle to follow a non-classical path due to self-interaction.

## 16. Particle Equation of Motion

The equation of motion for each particle can be obtained by using the usual Lagrange formula[25]:

$$\frac{d}{d\tau}\frac{\partial L}{\partial u^\alpha} - \frac{\partial L}{\partial x^\alpha} = 0 \tag{56}$$

This equation actually requires a Lagrangian L rather than a Lagrangian density $\mathcal{L}$. This can be obtained by re-expressing the overall Lagrangian density $\mathcal{L}$ in the form:

$$\mathcal{L} = \mathcal{L}_{field} + \sigma_0 L \tag{57}$$

in order to separate out the particle term L. Comparing the two different expressions (41) and (57) for $\mathcal{L}$, the Lagrangian for each particle's motion is seen to be:

$$L = \mp \rho_0 \left| u_\alpha u^\alpha \right|^{1/2} + u_\alpha j^\alpha \tag{58}$$

As shown in Appendix 5, substituting this expression into Eq. (56) yields the following equation of motion:

$$\frac{d(\rho_0 u_\alpha)}{d\tau} = \pm \frac{\partial \rho_0}{\partial x^\alpha} + u^\beta \left( \frac{\partial j_\alpha}{\partial x^\beta} - \frac{\partial j_\beta}{\partial x^\alpha} \right) \tag{59}$$

---

[25] See, e.g., [15].

where it is understood that the upper sign applies for timelike $u^\alpha$ and the lower sign applies for spacelike $u^\alpha$.

The form of this equation suggests that each particle moves as if it is under the influence of a scalar potential $\rho_0$ and a 4-vector potential $j^\alpha$. It is easily checked that this less restrictive equation is quite consistent with the equation of motion (55) introduced earlier via the joint distribution (54). Indeed, substituting expression (55) into Eq. (59) simply reduces the latter to an identity.

## 17. Field Equations

The proposed new Lagrangian density (41) will also generate field equations for the various wavefunctions $\psi_i$, $\psi_f$, $\psi_i^*$ and $\psi_f^*$. The wave equation for $\psi_i$, for example, can be found by varying $\psi_f^*$ in the Lagrangian density. This is achieved most simply by applying the usual Lagrangian formula[26], which here takes the form:

$$\partial_\alpha \frac{\partial \mathcal{L}}{\partial(\partial_\alpha \psi_f^*)} - \frac{\partial \mathcal{L}}{\partial \psi_f^*} = 0 \qquad (60)$$

The resulting wave equation will temporarily have a source term, which is to be expected in order to allow action and reaction between the particle and field and hence ensure conservation of energy and momentum. This source term will then be found to reduce to zero in the special limiting case of quantum mechanics as defined by the joint distribution (54).

Substituting the Lagrangian density expression (57) into Eq. (60), the following result is obtained:

$$\partial_\alpha \frac{\partial \mathcal{L}_{\text{field}}}{\partial(\partial_\alpha \psi_f^*)} - \frac{\partial \mathcal{L}_{\text{field}}}{\partial \psi_f^*} = -\left[\partial_\alpha \frac{\partial}{\partial(\partial_\alpha \psi_f^*)} - \frac{\partial}{\partial \psi_f^*}\right]\sigma_0 L \qquad (61)$$

which can be written equivalently as:

$$\partial_\alpha \frac{\partial \mathcal{L}_{\text{field}}}{\partial(\partial_\alpha \psi_f^*)} - \frac{\partial \mathcal{L}_{\text{field}}}{\partial \psi_f^*} = -\partial_\alpha \left[\sigma_0 \frac{\partial L}{\partial j^\beta} \frac{\partial j^\beta}{\partial(\partial_\alpha \psi_f^*)}\right] + \sigma_0 \frac{\partial L}{\partial j^\beta} \frac{\partial j^\beta}{\partial \psi_f^*} \qquad (62)$$

This is the general form of the wave equation for $\psi_i$. Taking the Dirac case of Sec. 14 as an example, the following equation for $\psi_i$ is then obtained:

$$i\gamma^\alpha \partial_\alpha \psi_i - m\psi_i = \sigma_0 \frac{\partial L}{\partial j^\alpha}\gamma^\alpha \psi_i \qquad (63)$$

---

[26] See, e.g., Sec. 7g in [13].

Here the left hand side of this equation contains the standard Dirac terms and the right hand side contains the new source term.

Once the velocity restriction implied by the joint distribution (54) is taken into account, the field equations are considerably simplified. In this regard, note that each source term on the right of both (62) and (63) contains a factor of the form $\frac{\partial L}{\partial j^\beta}$. As shown in Appendix 6, this factor is reduced to zero by Eq. (54) and so all the source terms also become zero. The Dirac equation then reduces to its standard form:

$$i\gamma^\alpha \partial_\alpha \psi_i - m\psi_i = 0 \qquad (64)$$

It is interesting to note that quantum mechanics in this model corresponds to an equilibrium situation where the source terms are unchanging and vanish.

The wave equations quoted so far in this section have only concerned the initial wavefunction $\psi_i$. A wave equation for the final wavefunction $\psi_f$, however, can easily be found as well. Varying $\psi_i^*$ will yield the wave equation for $\psi_f$ via the formula:

$$\partial_\alpha \frac{\partial \mathcal{L}}{\partial(\partial_\alpha \psi_i^*)} - \frac{\partial \mathcal{L}}{\partial \psi_i^*} = 0 \qquad (65)$$

After applying the velocity restriction (55), the wave equations for $\psi_f$ in both the Klein-Gordon and Dirac cases are found to be, respectively:

Klein-Gordon: $\qquad \partial_\alpha \partial^\alpha \psi_f + m^2 \psi_f = 0 \qquad (66)$

Dirac: $\qquad i\gamma^\alpha \partial_\alpha \psi_f - m\psi_f = 0 \qquad (67)$

which demonstrates in each case that the final wavefunction $\psi_f$ satisfies the same wave equation as the initial wavefunction $\psi_i$.

## 18. Conserved Current Density

In Secs. 13 and 14, where the Lagrangian densities for the Klein-Gordon and Dirac cases are presented as examples, the corresponding 4-current densities are simply stated without further justification in Eqs. (44) and (47). These are of the general form:

$$j^\alpha(x) = \mathrm{Re}\, \frac{\psi_f^*(x)\, \hat{j}^\alpha\, \psi_i(x)}{\langle f | i \rangle} \qquad (68)$$

which has been discussed at length in earlier sections. There is, however, a need to demonstrate that 4-current density expressions of this form can also be derived from the present Lagrangian density once Eq. (54) is assumed. To this end, the Klein-Gordon 4-current density will now be derived by way of example.

The usual assumption of global gauge invariance, when applied to the Lagrangian density (41), implies the existence of an associated conserved current density via Noether's theorem. In the present case, the relevant gauge transformation takes the form:

$$\begin{aligned} \psi_i &\to e^{i\phi}\psi_i \\ \psi_f &\to e^{i\phi}\psi_f \end{aligned} \qquad (69)$$

and leads to the following formula[27] for the 4-current density:

$$J^\alpha(x) = i\left[\frac{\partial \mathcal{L}}{\partial(\partial_\alpha \psi_i)}\psi_i + \frac{\partial \mathcal{L}}{\partial(\partial_\alpha \psi_f)}\psi_f - \frac{\partial \mathcal{L}}{\partial(\partial_\alpha \psi_i^*)}\psi_i^* - \frac{\partial \mathcal{L}}{\partial(\partial_\alpha \psi_f^*)}\psi_f^*\right] \qquad (70)$$

Substituting the Lagrangian expression (57) into Eq. (70) leads to the following expression for $J^\alpha(x)$:

$$\begin{aligned} J^\alpha(x) = &\; i\left[\frac{\partial \mathcal{L}_{\text{field}}}{\partial(\partial_\alpha \psi_i)}\psi_i + \frac{\partial \mathcal{L}_{\text{field}}}{\partial(\partial_\alpha \psi_f)}\psi_f - \frac{\partial \mathcal{L}_{\text{field}}}{\partial(\partial_\alpha \psi_i^*)}\psi_i^* - \frac{\partial \mathcal{L}_{\text{field}}}{\partial(\partial_\alpha \psi_f^*)}\psi_f^*\right] \\ &+ i\left[\frac{\partial(\sigma_0 L)}{\partial(\partial_\alpha \psi_i)}\psi_i + \frac{\partial(\sigma_0 L)}{\partial(\partial_\alpha \psi_f)}\psi_f - \frac{\partial(\sigma_0 L)}{\partial(\partial_\alpha \psi_i^*)}\psi_i^* - \frac{\partial(\sigma_0 L)}{\partial(\partial_\alpha \psi_f^*)}\psi_f^*\right] \end{aligned} \qquad (71)$$

which can be written equivalently as:

$$\begin{aligned} J^\alpha(x) = &\; i\left[\frac{\partial \mathcal{L}_{\text{field}}}{\partial(\partial_\alpha \psi_i)}\psi_i + \frac{\partial \mathcal{L}_{\text{field}}}{\partial(\partial_\alpha \psi_f)}\psi_f - \frac{\partial \mathcal{L}_{\text{field}}}{\partial(\partial_\alpha \psi_i^*)}\psi_i^* - \frac{\partial \mathcal{L}_{\text{field}}}{\partial(\partial_\alpha \psi_f^*)}\psi_f^*\right] \\ &+ i\sigma_0 \frac{\partial L}{\partial j^\beta}\left[\frac{\partial j^\beta}{\partial(\partial_\alpha \psi_i)}\psi_i + \frac{\partial j^\beta}{\partial(\partial_\alpha \psi_f)}\psi_f - \frac{\partial j^\beta}{\partial(\partial_\alpha \psi_i^*)}\psi_i^* - \frac{\partial j^\beta}{\partial(\partial_\alpha \psi_f^*)}\psi_f^*\right] \end{aligned} \qquad (72)$$

As usual, the velocity restriction implied by the joint distribution (54) now needs to be taken into account. Using the result derived in Appendix 6, the derivative $\frac{\partial L}{\partial j^\beta}$ in Eq. (72) becomes zero and so the 4-current density $J^\alpha(x)$ reduces to:

---

[27] See, e.g., Sec. 7g in [13]. An upper case J is employed here to distinguish the 4-current density expression now being considered from the usual $j^\alpha(x)$ discussed previously.

$$J^\alpha(x) = i\left[\frac{\partial \mathcal{L}_{field}}{\partial(\partial_\alpha \psi_i)}\psi_i + \frac{\partial \mathcal{L}_{field}}{\partial(\partial_\alpha \psi_f)}\psi_f - \frac{\partial \mathcal{L}_{field}}{\partial(\partial_\alpha \psi_i^*)}\psi_i^* - \frac{\partial \mathcal{L}_{field}}{\partial(\partial_\alpha \psi_f^*)}\psi_f^*\right] \quad (73)$$

For the Klein-Gordon case, substituting the $\mathcal{L}_{field}$ part of expression (43) into Eq. (73) then yields:

$$J^\alpha(x) = -\frac{1}{2m}\mathrm{Re}\,\frac{\psi_f^* i\overleftrightarrow{\partial}^\alpha \psi_i}{\langle f|i\rangle} \quad (74)$$

This is seen to be identical to the expression given in Eq. (44) and so the required result $J^\alpha(x) = j^\alpha(x)$ has thus been obtained.

This result also illustrates the wider point that the Lagrangian formalism presented in this paper generates 4-current density expressions of the general form (68). Note that these expressions are conserved via Noether's theorem and can easily be shown to satisfy the continuity equation, which was one of the requirements specified earlier in Sec. 2.

### 19. Energy-Momentum Tensor

Under the assumption that the Lagrangian density is not an explicit function of the coordinates $x^\alpha$ (i.e., that it is symmetric under space and time displacements), Noether's theorem implies the existence of an energy-momentum tensor $T^{\alpha\beta}$ for the particle-field system with this tensor having zero 4-divergence:

$$\partial_\beta T^{\alpha\beta} = 0 \quad (75)$$

This condition ensures overall conservation of energy and momentum. If the Lagrangian density is of the form (57):

$$\mathcal{L} = \mathcal{L}_{field} + \sigma_0 L \quad (76)$$

then, as shown in [16], the tensor $T^{\alpha\beta}$ is expressible naturally in the form of three terms:

$$T^{\alpha\beta} = T^{\alpha\beta}_{field} + T^{\alpha\beta}_{particle} + T^{\alpha\beta}_{interaction} \quad (77)$$

where the individual terms are given by[28]:

---

[28] The overall tensor $T^{\mu\nu}$ defined here is actually the "canonical" energy-momentum tensor, which is not necessarily symmetric and hence does not necessarily conserve angular momentum. Techniques exist to symmetrise this tensor [17].

$$T^{\alpha\beta}_{field} = (\partial^{\alpha}\psi_i)\frac{\partial \mathcal{L}_{field}}{\partial(\partial_{\beta}\psi_i)} + (\partial^{\alpha}\psi_f)\frac{\partial \mathcal{L}_{field}}{\partial(\partial_{\beta}\psi_f)} + (\partial^{\alpha}\psi_i^*)\frac{\partial \mathcal{L}_{field}}{\partial(\partial_{\beta}\psi_i^*)} + (\partial^{\alpha}\psi_f^*)\frac{\partial \mathcal{L}_{field}}{\partial(\partial_{\beta}\psi_f^*)} - g^{\alpha\beta}\mathcal{L}_{field}$$
(78)

$$T^{\alpha\beta}_{particle} = \sigma_0 p^{\alpha} u^{\beta}$$
(79)

$$T^{\alpha\beta}_{interaction} = \sigma_0(\partial^{\alpha}\psi_i)\frac{\partial L}{\partial(\partial_{\beta}\psi_i)} + \sigma_0(\partial^{\alpha}\psi_f)\frac{\partial L}{\partial(\partial_{\beta}\psi_f)} + \sigma_0(\partial^{\alpha}\psi_i^*)\frac{\partial L}{\partial(\partial_{\beta}\psi_i^*)} + \sigma_0(\partial^{\alpha}\psi_f^*)\frac{\partial L}{\partial(\partial_{\beta}\psi_f^*)}$$
(80)

In the particle term, the quantity $p^{\alpha}$ is the particle's generalised 4-momentum [15]:

$$p^{\alpha} \equiv -\frac{\partial L}{\partial u_{\alpha}}$$
(81)

For the many-particle case discussed in this paper, each of the three quantities $T^{\alpha\beta}_{field}$, $T^{\alpha\beta}_{particle}$ and $T^{\alpha\beta}_{interaction}$ will be the sum of separate contributions from the individual particles.

For the present Lagrangian, the term $T^{\alpha\beta}_{interaction}$ can be written more specifically as:

$$T^{\alpha\beta}_{interaction} = \sigma_0 \frac{\partial L}{\partial j^{\lambda}}\left[(\partial^{\alpha}\psi_i)\frac{\partial j^{\lambda}}{\partial(\partial_{\beta}\psi_i)} + (\partial^{\alpha}\psi_f)\frac{\partial j^{\lambda}}{\partial(\partial_{\beta}\psi_f)} + (\partial^{\alpha}\psi_i^*)\frac{\partial j^{\lambda}}{\partial(\partial_{\beta}\psi_i^*)} + (\partial^{\alpha}\psi_f^*)\frac{\partial j^{\lambda}}{\partial(\partial_{\beta}\psi_f^*)}\right]$$
(82)

Now, once the joint distribution (54) is taken into account as usual, considerable simplification is introduced. In particular, the interaction term vanishes (using the result in Appendix 6) and the divergences of $T^{\alpha\beta}_{field}$ and $T^{\alpha\beta}_{particle}$ become separately zero, indicating that there is no interchange of energy and momentum between the field and the particle. Furthermore, in Eq. (79) the particle's generalised 4-momentum:

$$p^{\alpha} = -\frac{\partial L}{\partial u_{\alpha}} = \rho_0 u^{\alpha} - j^{\alpha} \qquad \text{(see Eq. (106))}$$
(83)

becomes zero. As well as reducing $T^{\alpha\beta}_{particle}$ to zero, this also means that $p^{\alpha}$ has become constant, in accordance with conservation of the particle's energy and momentum.

In general, the expression for $T^{\alpha\beta}_{field}$ for each particle will be of the form[29]:

---

[29] This expression has been suggested before in Secs. 9 and 11 of [18], and in [19].

$$T^{\alpha\beta}_{\text{field}} = \text{Re}\frac{\psi_f^* \hat{T}^{\alpha\beta} \psi_i}{\langle f|i\rangle} \tag{84}$$

where $\hat{T}^{\alpha\beta}$ is the energy-momentum operator provided by the particular wave equation under consideration. By analogy with the simple proof in Sec. 5, this expression can be shown to reduce to the standard expression $T^{\alpha\beta}_{\text{field}} = \psi_i^* \hat{T}^{\alpha\beta} \psi_i$ when only the initial state is known and a weighted average is taken over the final states.

Note that the energy-momentum tensor (84) is not statistical. Given the initial and final boundary conditions $|i\rangle$ and $|f\rangle$, the tensor is uniquely determined. In contrast, the particle positions are distributed statistically. Specifying both $|i\rangle$ and $|f\rangle$ uniquely determines the 4-current density $j^{\alpha}(x)$ but leaves the particle positions spread in accordance with the distribution given by the zeroth component $j^0(x)$.

## 20. Discussion and Conclusions

The aim here of formulating a Lagrangian description for a particle interpretation of quantum mechanics has been successfully carried out for the general case of entangled many-particle states under the assumption that the particles have ceased interacting[30]. The model provides a separate Lagrangian density for each particle in four-dimensional spacetime, thereby avoiding the need to resort to configuration space. Both the availability of a spacetime description and the ability to maintain Lorentz invariance are made possible by incorporating final boundary conditions and retrocausality into the model. The proposed Lagrangian density expression then provides all the usual formalism for answering any question we wish to ask. It provides a clear picture of events at all times, accompanied by field equations, particle equations of motion and conservation of energy and momentum.

The Lagrangian expression nominated in this model seems to be the only one that is consistent with a particle ontology for relativistic quantum mechanics in the many-particle case. Within the non-statistical version of the model, where the Lagrangian density is not accompanied by the joint distribution assumption (54), the particle acts as the source of a field and the particle and field then mutually interact via the particle equation of motion (59) and the relevant field equation, e.g., Eq. (63). However, once the statistical assumption (54) is also included, we obtain the special case corresponding to quantum mechanics. The particle equation of motion then simplifies to the form of a guidance equation and the source term in the field equation becomes zero, resulting in a retrocausal version of the Bohm model. This is essentially the same as the "causally symmetric Bohm model" formulated in [10], the only difference being that the earlier presentation was mainly non-relativistic, whereas now the formulation is fully Lorentz invariant (as well as being extended to Lagrangian form). This causally symmetric model reduces back further to the standard Bohm model once a weighted average is taken over the unknown future boundary conditions and they are integrated out.

---

[30] A possible method of generalising this model to the cases of continuing interaction and of creation and annihilation is outlined in Secs. 8 and 9 of [18].

In narrowing to the quantum mechanical case via assumption (54) it should be noted that, although the source term goes to zero, this does not mean that the field becomes zero. The field is still there propagating with the corresponding particle and influencing the particle's trajectory (as in the standard Bohm model) but it is not actually being emitted or absorbed by the particle. If the particle's velocity were to stray away from the value enforced by Eq. (54), there would then be net emission or absorption. In this context, the existence of the field could perhaps be attributed to some "non-equilibrium" period in the past or future when the particle's 4-velocity was/will be different from that given by (54). In such a non-equilibrium state, the Born probability rule would also no longer hold, thereby introducing the possibility of new experimental consequences (as has been pointed out previously in the context of the standard Bohm model [20-22]). An interesting topic for further research would be to explore whether the equation of motion (59) presented here tends to restore each particle to the equilibrium state corresponding to quantum mechanics, thereby providing an explanation for the persistence of this special case. Note that the particle's generalised 4-momentum is conserved in this case and so no energy or momentum is exchanged with the field. This does not mean, however, that the particle's 4-velocity is constant, as can be seen from expression (83) for the generalised 4-momentum derivable from the Lagrangian density. This definition allows the 4-velocity to vary continuously with position via Eq. (55) in such a way that consistency with the quantum mechanical predictions is maintained.

It should also be noted that taking the present step of adding retrocausality into the standard Bohm model introduces a number of improvements into that model. In particular:

1. The model can easily be set in Lorentz invariant form

2. The model becomes local from a spacetime viewpoint

3. A general form of the model can be formulated which is applicable for any wave equation

4. A configuration space ontology is avoided, with the many-particle case remaining in four-dimensional spacetime

5. A separate velocity expression can be provided for each of the n particles in an entangled state, rather than just a single, overall velocity defined in 3n dimensions

6. The correct statistical correlations can be maintained while employing a separate wavefunction for each particle

7. A physical interpretation can be provided for the negative values of the Klein-Gordon "probability" density [23]

8. A Lagrangian formulation becomes possible

9. Energy and momentum conservation is restored

10. A source is provided for the guiding field in the more general case

11. Action-reaction between the particle and the field is restored in this case.

Concerning possible future work, although the configuration-space wavefunction for many particles has been successfully excluded from the ontology here, it remains a relevant mathematical function in this model. A more ambitious project for the future might be a formulation expressed in terms of the ontological quantities only and not involving this wavefunction. Finally, a possible application of this model is that it opens an alternative path to quantum gravity by providing a definite, ontological energy-momentum tensor for insertion in Einstein's gravitational field equation. This non-statistical tensor thereby allows the Einstein curvature tensor on the other side of the equation to remain non-statistical and unquantised without introducing any mismatch or inconsistency. A more detailed formulation of this approach can be found in [19].

**Acknowledgements** The author would especially like to thank Ken Wharton for many insightful discussions, plus David Miller, Peter Evans and Dustin Lazarovici for their helpful feedback on this work.

**Appendix 1: Relativistic Entangled State for a Pair of Particles**

This section expands on the discussion in Sec. 3. The initial state (i.e., the initial boundary condition) in any physical theory is commonly expressed on a chosen hyperplane of constant time $t_i$. Hence, for the entangled pair of particles considered in Eq. (2), the initial state is expected to be of the form $\langle \mathbf{x}_i, \mathbf{x}'_i; t_i | i \rangle$. For later times, a state of the form $\langle x, x' | i \rangle \equiv \langle \mathbf{x}, t; \mathbf{x}', t' | i \rangle$ can then be obtained from this initial one via the relationship:

$$\langle \mathbf{x}, t; \mathbf{x}', t' | i \rangle = \int_{-\infty}^{+\infty} \int_{-\infty}^{+\infty} K(\mathbf{x}, t; \mathbf{x}_i, t_i) \hat{j}^0 K(\mathbf{x}', t'; \mathbf{x}'_i, t_i) \hat{j}^{0'} \langle \mathbf{x}_i, \mathbf{x}'_i; t_i | i \rangle d^3 x_i \, d^3 x'_i \tag{85}$$

where the K's are propagators which are solutions of the relevant wave equation and carry the state forwards in time from $t_i$ to $t$ and $t'$. The resulting multi-time entangled state provides the joint probability density $|\langle x, x' | i \rangle|^2$ for the 1st particle to be at position $\mathbf{x}$ at time $t$ and the 2nd particle to be at $\mathbf{x}'$ at time $t'$. The single-particle state $\langle x' | i \rangle$ defined in Eq. (2) is then seen to be the projection generated by the measurement on the 1st particle and can be identified as the 2nd particle's resulting new state.

Note that by inserting Eq. (85) into Eq. (2) and using the standard relationship:

$$\int_{-\infty}^{+\infty} \langle f | \mathbf{x}, t \rangle \hat{j}^0 K(\mathbf{x}, t; \mathbf{x}_i, t_i) d^3 x = \langle f | \mathbf{x}_i, t_i \rangle \tag{86}$$

which propagates the state $\langle f|\mathbf{x},t\rangle$ back to time $t_i$, the following result is obtained:

$$\langle x'|i\rangle = \frac{1}{N}\int_{-\infty}^{+\infty}\int_{-\infty}^{+\infty} K(\mathbf{x}',t';\mathbf{x}'_i,t_i)\,\hat{j}^{0'}\,\langle f|\mathbf{x}_i,t_i\rangle\,\hat{j}^{0}\,\langle\mathbf{x}_i,\mathbf{x}'_i;t_i|i\rangle\,d^3x_i\,d^3x'_i \qquad (87)$$

which shows that the updated state $\langle x'|i\rangle$ for the 2nd particle is not a function of the time parameter t associated with the 1st particle's state.

**Appendix 2: Position Measurement as the Final Boundary Condition**

Here it will be shown that, for the special case where the next measurement performed is one of position, the two requirements listed at the end of Sec. 6 are satisfied. The proof will be presented in nonrelativistic form (see footnote 13).

For the Schrodinger case, the time component of Eq. (8) reduces to:

$$j^0(x) = \mathrm{Re}\,\frac{\langle f|x\rangle\langle x|i\rangle}{\langle f|i\rangle} \qquad (88)$$

Taking the result of the final position measurement to be $x_f$, this equation then becomes:

$$j^0(x) = \mathrm{Re}\,\frac{\langle x_f|x\rangle\langle x|i\rangle}{\langle f|i\rangle} \qquad (89)$$

It can then be written out with all relevant times shown explicitly as follows:

$$j^0(\mathbf{x},t) = \mathrm{Re}\,\frac{\langle \mathbf{x}_f,t_f|\mathbf{x},t\rangle\langle \mathbf{x},t|i\rangle}{\langle \mathbf{x}_f,t_f|i\rangle} \qquad (90)$$

In approaching the limit $t\to t_f$, it is a standard result that the term $\langle \mathbf{x}_f,t_f|\mathbf{x},t\rangle$ gradually becomes a delta function:

$$\langle \mathbf{x}_f,t_f|\mathbf{x},t\rangle \to \delta^3(\mathbf{x}-\mathbf{x}_f) \qquad (91)$$

Hence in this limit Eq. (90) becomes:

$$j^0(\mathbf{x},t_f) = \mathrm{Re}\,\frac{\delta^3(\mathbf{x}-\mathbf{x}_f)\langle \mathbf{x},t_f|i\rangle}{\langle \mathbf{x}_f,t_f|i\rangle} \qquad (92)$$

The delta function then allows cancellation to yield:

$$j^0(\mathbf{x},t_f) = \delta^3(\mathbf{x}-\mathbf{x}_f) \qquad (93)$$

This result shows, as required, that the time component of the 4-current density has become positive in approaching the measurement time and that it has become zero for all position values other than the one measured.

**Appendix 3: Transluminal World Lines**

It is a feature of the present model that it accepts and employs world lines which continue curving until they pass smoothly through the light barrier (in the manner of the current lines corresponding to the Klein-Gordon 4-current density). This may be an unorthodox notion but is, in fact, fully compatible with special relativity under certain circumstances. It is simply a matter of such paths not usually being taken seriously because they are not observed experimentally. In this regard, the theory presented here automatically restricts such behaviour to times between observations, so the absence of evidence is easily resolved. There is, however, an extra condition which needs to be incorporated into the model in order for such a generalisation to be mathematically manageable, namely that the particle's mass should also drop smoothly to zero as the light barrier is approached.

The technique for describing such "transluminal" world lines is then to work with 4-momentum (equal to mass times 4-velocity), rather than with mass and 4-velocity separately. Then, even though the 4-velocity goes to infinity as the light barrier is approached, the mass goes to zero in such a way that the product is well-behaved and remains finite and differentiable throughout the transition.

This is illustrated by the equation of motion (59), where the left hand side contains the rate of change of the product $\rho_0 u_\alpha$. Here $\rho_0$ plays the role of mass and goes to zero as $u_\alpha$ goes to infinity such that the product (essentially the 4-momentum[31]) remains finite. This finiteness can be verified from the fact that, once the joint distribution (54) is introduced, the product $\rho_0 u_\alpha$ is always equal to the finite quantity $j_\alpha$.

**Appendix 4: Mathematical Treatment of Spacelike Segments**

As already discussed in Sec. 6 and Appendix 3, the proposed current density 4-vector defined by Eq. (8) is not limited to timelike directions in spacetime and can point in spacelike directions as well. To handle this situation mathematically, there is a need here to extend the usual relationships for proper time and 4-velocity to the spacelike domain.

The 4-velocity of a particle:

$$u^\alpha = \frac{dx^\alpha}{d\tau} \tag{94}$$

---

[31] Strictly speaking a constant needs to be included here in order to give the correct dimensions for 4-momentum, this being the same constant mentioned in footnote 17.

is defined in terms of the proper time $\tau$, which is taken to be a variable which increases monotonically as we go along the particle's world line. The proper time simply marks off distance along the line in four dimensions. It is clear then that $\tau$ needs to be always real. Taking the metric signature to be $(+---)$, this means that the usual definition:

$$d\tau = \left(dt^2 - dx^2 - dy^2 - dz^2\right)^{1/2}$$
$$\equiv \left(dx_\alpha dx^\alpha\right)^{1/2} \tag{95}$$

which applies for a change $d\tau$ along a time-like segment of the world line, must be supplemented with the definition:

$$d\tau = \left(-dx_\alpha dx^\alpha\right)^{1/2} \tag{96}$$

for the case of a space-like segment in order to keep $\tau$ real. This two-part definition for proper time is relativistically invariant because all observers agree as to whether any given segment is time-like or space-like. The definition can be written as:

$$d\tau = \begin{cases} \left(dx_\alpha dx^\alpha\right)^{1/2} & \text{(time–like segment)} \\ \left(-dx_\alpha dx^\alpha\right)^{1/2} & \text{(space–like segment)} \end{cases} \tag{97}$$

or, if preferred, summarized in the single expression [12]:

$$d\tau = \left|dx_\alpha dx^\alpha\right|^{1/2} \tag{98}$$

Now Eq. (94) implies the following relationship:

$$u_\alpha u^\alpha = \frac{dx_\alpha dx^\alpha}{(d\tau)^2} \tag{99}$$

Combining this with Eq. (97), it is seen that any time-like 4-velocity vector will satisfy the identity:

$$u_\alpha u^\alpha = 1 \tag{100}$$

whereas any space-like 4-velocity will satisfy:

$$u_\alpha u^\alpha = -1 \tag{101}$$

with the following identity holding for any 4-velocity vector:

$$\left|u_\alpha u^\alpha\right| = 1 \tag{102}$$

Similarly, the current density 4-vector $j^\alpha = \rho_0 \bar{u}^\alpha$ as defined in Eq. (35) will satisfy:

$$(j_\alpha j^\alpha)^{1/2} = \rho_0 \tag{103}$$

when this 4-vector is timelike, whereas it will satisfy:

$$(-j_\alpha j^\alpha)^{1/2} = \rho_0 \tag{104}$$

when it is spacelike. The following identity will hold in either case:

$$\left| j_\alpha j^\alpha \right|^{1/2} = \rho_0 \tag{105}$$

**Appendix 5: Particle Equation of Motion**

The aim here is to derive the equation of motion for a particle, as required in Sec. 16. To this end, it will be helpful first to derive a simple identity for the derivative $\dfrac{\partial L}{\partial u^\alpha}$, since it will be needed both here and in Sec. 19. The following proof assumes the convention that an upper sign refers to the case where $u^\alpha$ is timelike, whilst a lower sign applies for spacelike $u^\alpha$. From Eq. (58), the required derivative is then found to be:

$$\begin{aligned}
\frac{\partial L}{\partial u^\alpha} &= \frac{\partial \left[ \mp \rho_0 \left( \pm u_\beta u^\beta \right)^{1/2} + u_\beta j^\beta \right]}{\partial u^\alpha} \quad \text{(using Eqs. (100) and (101))} \\
&= \mp \rho_0 \tfrac{1}{2} \left( \pm u_\beta u^\beta \right)^{-1/2} \left( \pm g_{\alpha\lambda} u^\lambda \pm u_\lambda \delta^\lambda_\alpha \right) + g_{\alpha\beta} j^\beta \\
&= \mp \rho_0 \tfrac{1}{2} (1)(\pm 2 u_\alpha) + j_\alpha \\
&= j_\alpha - \rho_0 u_\alpha
\end{aligned}$$

$$\tag{106}$$

Substituting this result together with the Lagrangian (58) into the left hand side of the Lagrange formula (56), the following is then obtained:

$$\begin{aligned}
\frac{d}{d\tau} \frac{\partial L}{\partial u^\alpha} - \frac{\partial L}{\partial x^\alpha} &= \frac{d}{d\tau}(j_\alpha - \rho_0 u_\alpha) - \frac{\partial \left( \mp \rho_0 \left| u_\beta u^\beta \right|^{1/2} + u_\beta j^\beta \right)}{\partial x^\alpha} \\
&= u^\beta \frac{\partial j_\alpha}{\partial x^\beta} - \frac{d(\rho_0 u_\alpha)}{d\tau} \pm \frac{\partial \rho_0}{\partial x^\alpha} - u_\beta \frac{\partial j^\beta}{\partial x^\alpha} \\
&= -\frac{d(\rho_0 u_\alpha)}{d\tau} \pm \frac{\partial \rho_0}{\partial x^\alpha} + u^\beta \left( \frac{\partial j_\alpha}{\partial x^\beta} - \frac{\partial j_\beta}{\partial x^\alpha} \right)
\end{aligned}$$

$$\tag{107}$$

Combining (107) and (56) finally yields:

$$\frac{d(\rho_0 u_\alpha)}{d\tau} = \pm \frac{\partial \rho_0}{\partial x^\alpha} + u^\beta \left( \frac{\partial j_\alpha}{\partial x^\beta} - \frac{\partial j_\beta}{\partial x^\alpha} \right) \tag{108}$$

with $+$ for timelike $u^\alpha$ and $-$ for spacelike $u^\alpha$.

**Appendix 6: Useful Identity**

A simple identity for the derivative $\frac{\partial L}{\partial j^\beta}$ will be derived here for use in Secs. 17 to 19. From Eq. (58), the relevant expression for L is:

$$\begin{aligned} L &= \mp \rho_0 + u_\alpha j^\alpha & \text{(using Eq. (102))} \\ &= -u_\alpha u^\alpha \rho_0 + u_\alpha j^\alpha & \text{(using Eqs. (100) and (101))} \\ &= -u_\alpha u^\alpha \left( \pm j_\nu j^\nu \right)^{1/2} + u_\alpha j^\alpha & \text{(using Eqs. (103) and (104))} \end{aligned} \tag{109}$$

where, in the last line, the upper sign applies for timelike $j^\alpha$ and the lower sign applies for spacelike $j^\alpha$. Hence the required derivative is:

$$\begin{aligned} \frac{\partial L}{\partial j^\beta} &= -u_\alpha u^\alpha \tfrac{1}{2} \left( \pm j_\nu j^\nu \right)^{-1/2} \left( \pm g_{\lambda\beta} j^\lambda \pm j_\lambda \delta^\lambda_\beta \right) + u_\alpha \delta^\alpha_\beta \\ &= -u_\alpha u^\alpha \tfrac{1}{2} \frac{1}{\rho_0} \left( \pm 2 j_\beta \right) + u_\beta \\ &= u_\beta \mp u_\alpha u^\alpha \frac{j_\beta}{\rho_0} \end{aligned} \tag{110}$$

This expression can be simplified by temporarily adopting a new sign convention that the upper sign applies when $u^\alpha$ and $j^\alpha$ are both timelike or both spacelike and the lower sign applies when $u^\alpha$ and $j^\alpha$ are different. Eq. (110) then becomes:

$$\frac{\partial L}{\partial j^\beta} = u_\beta \mp \frac{j_\beta}{\rho_0} \tag{111}$$

Now, once the velocity restriction implied by the joint distribution (54) is taken into account, the 4-vectors for $u^\alpha$ and $j^\alpha$ must be in the same direction in spacetime and so only the upper sign applies in (111). Using the restriction (54) a second time, the right hand side of (111) then vanishes yielding:

$$\frac{\partial L}{\partial j^\beta} = 0 \tag{112}$$

This is the result needed for simplifying the expressions in Secs. 17 to 19.